# The Uncertainty Principle derived by the finite transmission of light and information

**Piero Chiarelli**[1*]

[1] *National Council of Research of Italy, Area of Pisa, 56124 Pisa, Moruzzi 1, Italy*

## ABSTRACT

This work shows that in the frame of the stochastic generalization of the quantum hydrodynamic analogy (QHA) the uncertainty principle can be derived by the postulate of finite transmission speed of light and information . The theory shows that the measurement process performed in the large scale classical limit of stochastic QHA (SQHA), cannot have a duration smaller than the time need to the light to travel the distance up to which the quantum non-local interaction extend itself. The product of the minimum measuring time multiplied by the variance of energy fluctuation due to presence of stochastic noise shows to lead to the minimum uncertainty principle. The paper also shows that the uncertainty relations can be also derived if applied to the indetermination of position and momentum of a particle of mass m in a quantum fluctuating environment.

## 1. INTRODUCTION

How the classical behavior is achieved in passing from the quantum dynamics to the macroscopic scale is a problem of interest in many branches of physics [1]. Actually, a unified theory hosting both the quantum mechanics as well as the classical ones, where the transition between the two dynamics is gradual, coherent and systematic, is still underway [2-4].

The incompatibility between the quantum and classical mechanics comes mainly from the impossibility to manage the non-local interaction of the quantum mechanics that acts among the parts of a system whatever is their relative distance even if infinite. This infinite range of non-locality of the quantum mechanics makes it incompatible with the (local) classical mechanics that cannot be achieved on whatever large scale limit of quantum theory.

This missing bridge between the classical and the quantum mechanics leads also to the logical incompatibility between the quantum non-local interaction and the finite speed of transmission of light and information at the base of the relativistic theory.

* Tel.: +39-050-315-2359; fax: +39-050-315-2166.
E-mail address: pchiare@ifc.cnr.it.

The difficulty in passing from the quantum to classical description has a clear mathematical evidence in the QHA approach [5-7] where the classical limit is recovered simply by posing the Plank's constant $\hbar$ equal to zero.

Even if this procedure is well understandable from the empirical point of view, it is not correct from the mathematical standpoint. This because by posing $\hbar = 0$ and eliminating the quantum pseudo-potential [6] the stationary QHA distributions (i.e., the eigenfunctions of the Schrödinger approach) are wiped out so that the nature of the QHA equations are deeply changed.

Even if this is a great theoretical problem, from the empirical point of view, the solution appears quite natural. It has been shown by many authors that fluctuations may destroy quantum coherence and elicit the emergence of the classical behavior [8-10].

Large scale dynamics (i.e., $\hbar$ small and negligible quantum potential interaction) and fluctuations clearly seem to be the "ingredients" that can lead to the classical behavior.

This observation is confirmed by the stochastic generalization of the QHA developed by the author [11-12] where the mechanism to pass from the quantum to the classical behavior is analytically detailed .

In the SQHA model the standard quantum mechanics is obtained as the deterministic limit of the theory while the stochastic classical behavior is recovered in weakly interacting system as the large scale limit.

By using this model, where it is possible to systematically pass from the quantum behavior to the classical (stochastic) one, we show in this work that when a statistical measurement is performed (i.e., the measuring apparatus and system are not a unique quantum system) such a process has a finite minimum time of realization. This result is due to the fact that the measuring apparatus and the system must be far apart a distance bigger than the distance of quantum non-local interaction and the transmission of information cannot be faster than the light speed. The theory shows that the minimum interval of time measurement multiplied by the energy variance due to the environment fluctuations is a constant and that can be set equal to the Plank's constant.

## 2. THE SQHA EQUATION OF MOTION

When space distributed noise is considered in the QHA, the motion equation for the particles density (PD) n (i.e., the wave function modulus squared (WFMS)), in the limit of small noise amplitude reads [12]

$$\partial_t \mathrm{n}_{(q,t)} = -\nabla_q \bullet (\mathrm{n}_{(q,t)} \dot{q}) + \eta_{(q,t,\Theta)} \quad (1)$$

$$\lim_{\Theta \to 0} < \eta_{(q_\alpha,t)}, \eta_{(q_\beta+\lambda,t+\tau)} > = k\Theta \frac{\mu}{2\lambda_c^2} exp[-(\frac{\lambda}{\lambda_c})^2] \delta(\tau) \delta_{\alpha\beta} \quad (2)$$

where $\dot{q}_{(t)}$ is obtained by the solution of the differential equations

$$\dot{q} = \frac{p}{m}, \quad (4)$$

$$\dot{p} = -\nabla (V_{(q)} + V_{qu(\mathrm{n})}), \quad (5)$$

where the noise correlation distance $\lambda_c$ reads

$$\lim_{\Theta \to 0} \lambda_c = \pi \frac{\hbar}{(2mk\Theta)^{1/2}} \quad , \quad (3)$$

where $\Theta$ is the amplitude of the spatially distributed noise $\eta$, $\mu$ is the PD mobility that depends by the specificity of the considered system [12], $V_{(q)}$ represents the Hamiltonian potential and $V_{qu(n)}$ is the so-called (non-local) quantum potential [5-7] that reads

$$V_{qu} = -(\frac{\hbar^2}{2m})n^{-1/2}\nabla \bullet \nabla n^{1/2} . \qquad (6)$$

The Schrödinger equation is recovered for the complex variable [5]

$$\psi_{(q,t)} = n^{\frac{1}{2}}_{(q,t)} \exp[\frac{i}{\hbar} S_{(q,t)}] \qquad (7)$$

where $\nabla S_{(q,t)} = p$.

The noise variance (2) is a direct consequence of the derivatives present into the quantum potential (i.e., $\nabla \bullet \nabla n^{1/2}$) that gives rise to a membrane elastic-like contribution to system energy [5] that reads

$$\overline{H}_{qu} = \int_{-\infty}^{\infty} n_{(q,t)} V_{qu(q,t)} dq , \qquad (8)$$

where higher curvature of the PD $n^{1/2}$ leads to higher energy (see figure 1).

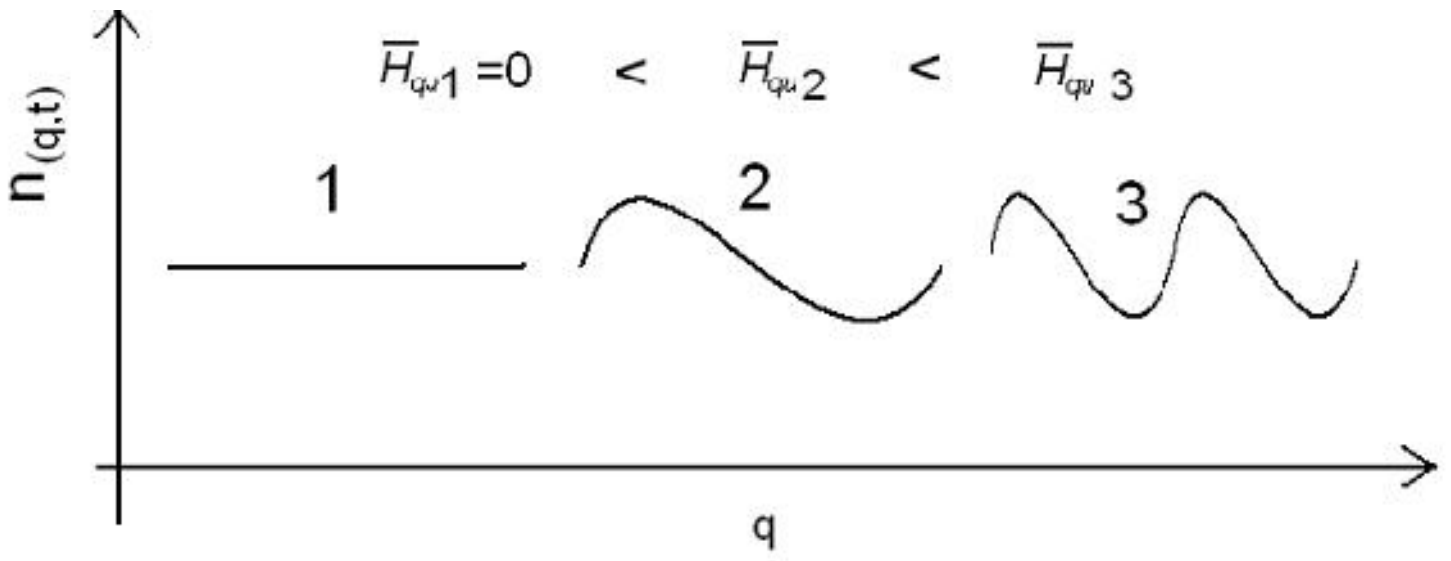

Figure 1 Quantum potential energy as a function of the "curvature" of the particle density n.

Therefore, independent fluctuations infinitesimally far apart and leading to very large curvature wrinkles of the PD and hence to an infinite quantum potential energy.

If we require that any physical system can attain configurations with finite energy, independent fluctuations of PD on shorter and shorter distance must have smaller and smaller amplitude. In the small noise limit, the energetic constraint

$$\lim_{\lambda_c \to 0} \overline{H}_{qu} = const < \infty$$

leads to the existence of a correlation distance (let's name it $\lambda_c$) for the noise variance [12] (see figure2) By imposing that the variance of the quantum potential energy fluctuations does not diverge in presence of a Gaussian noise it follows that the correlation distance must obey to the relation [12]

$$\lim_{\Theta \to 0} \lambda_c \propto \frac{\hbar}{(2mk\Theta)^{1/2}} . \qquad (9)$$

The proportionality constant is defined by additional information that we discuss about further on in the paper.

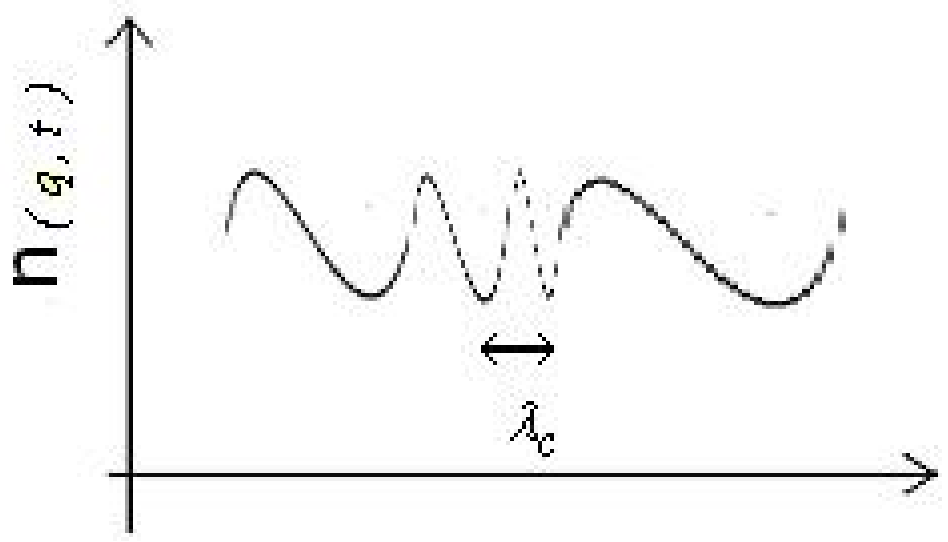

Figure 2. The particle density $n_{(q,t)}$ owing Gaussian Fluctuations

It is worth mentioning that the stochastic equations (1-5) deriving by the following system of differential equations

$$\partial_t S_{(q,t)} = -V_{(q)} + \frac{\hbar^2}{2m}\frac{\nabla^2 A_{(q,t)}}{A_{(q,t)}} - \frac{1}{2m}(\nabla S_{(q,t)})^2 \qquad (10)$$

$$\partial_t A_{(q,t)} = -\frac{1}{m}\nabla A_{(q,t)} \bullet \nabla S_{(q,t)} - \frac{1}{2m}A\nabla^2 S_{(q,t)} + A^{-1}\eta_{(q_\alpha,t,\Theta)} \qquad (11)$$

for the complex variable (7), are equivalent to the stochastic Schrödinger equation [11]

$$i\hbar\frac{\partial\psi}{\partial t} = -\frac{\hbar^2}{2m}\nabla^2\psi + V_{(q)}\psi + i\frac{\psi}{|\psi|^2}\eta_{(q_\alpha,t,\Theta)} \cdot \qquad (12)$$

## 2.1 Local large-scale dynamics

In addition to the noise correlation function (2), in the large-distance limit, it is also important to know the behavior of the quantum force $\dot{p}_{qu} = -\nabla_q V_{qu}$.

The relevance of the force generated by the quantum potential at large distance can be evaluated by the convergence of the integral [2]

$$\int_0^\infty | q^{-1}\nabla V_{qu} | dq \qquad (13)$$

If the quantum potential force $-\nabla_q V_{qu}$ goes like $\lim_{|q|\to\infty} |\nabla V_{qu}| \propto | q |^{-\varepsilon}$, where $\varepsilon > 0$, at large distance so that $\lim_{|q|\to\infty} | q^{-1}\nabla V_{qu} | \propto | q |^{-(1+\varepsilon)}$ the integral (13) converges. In this case, the mean weighted distance

$$\lambda_q = 2\frac{\int_0^\infty | q^{-1}\frac{dV_{qu}}{dq} | dq}{\lambda_c^{-1} | \frac{dV_{qu}}{dq} |_{(q=\lambda_c)}}, \qquad (14)$$

can be used to evaluate the quantum potential range of interaction.

For analyzing the interplay between the Hamiltonian potential and the quantum one in a general non-linear case, we can consider the linear case. For the linear interaction, the Gaussian-type eigenstates leads to a quadratic quantum potential and, hence, to a linear quantum force, so that $\lim_{|q|\to\infty} | q^{-1}\nabla V_{qu} | \propto constant$ and $\lambda_q$ diverges.

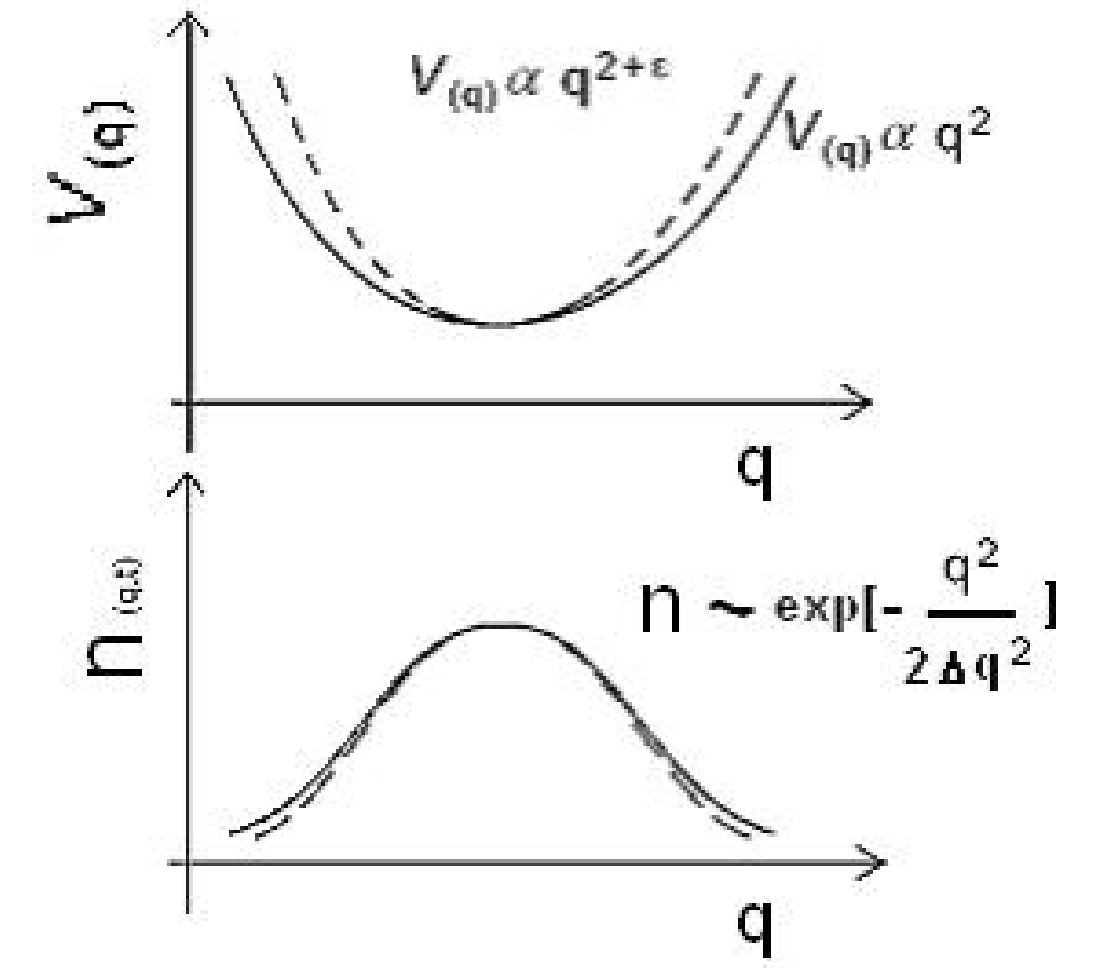

Figure 3. Particle density $n_{(q,t)}$ for near-linear Hamiltonian potential

Moreover, since faster the Hamiltonian potential grows, more localized is the WFMS and hence stronger is the quantum potential, in order to have $\lambda_q$ finite (so that the large-scale classical limit can be achieved) we have to deal with a system of particles interacting by a

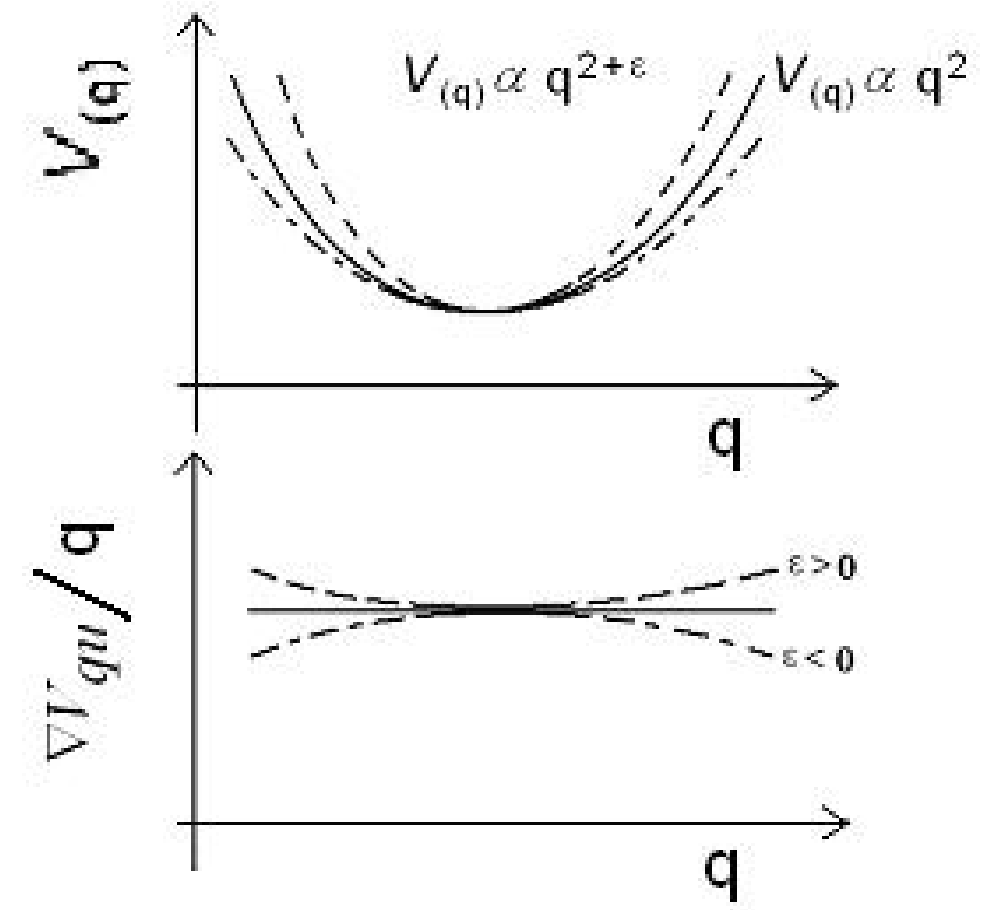

Figure4. The scaled quantum force $q^{-1}\nabla V_{qu}$ for near-linear Hamiltonian potential weaker than the linear interaction.

For Hamiltonian potential weaker that the quadratic one ($\varepsilon<0$ in figure 4), the quantum potential force (QPF) $\nabla V_{qu}$ at large distance grows less than linearly. Given the case of a pseudo-Gaussian WFM such as $\lim_{|q|\to\infty} n \propto exp[-P^k(q)]$, for $k < \frac{2}{3}$ the integral $\int_0^\infty | q^{-1} \nabla V_{qu} | dq$ converges [12].

When the length resolution $\Delta\Omega$ of our physical problem is larger than $\lambda_c$ and $\lambda_q$ so that we have $\lambda_c \cup \lambda_q << \Delta\Omega << \Delta L$, where $\Delta L$ is the physical length of the system, equation (11) reads [12]

$$\lim_{\Theta\to 0} \dot{q} = \frac{p}{m} = \lim_{\Delta L/\lambda_c\to\infty} \lim_{\Delta L/\lambda_q\to\infty} \lim_{\Theta\to 0} \frac{\nabla S}{m}$$

$$= \nabla\{ \lim_{\Delta L/\lambda_c\to\infty} \lim_{\Delta L/\lambda_q\to\infty} \lim_{\Theta\to 0} \frac{1}{m} \int_{t_0}^{t} dt( \frac{p\bullet p}{2m} - V_{(q)} - V_{qu(\mathrm{n})} )\}$$

$$= \nabla\{ \lim_{\Delta L/\lambda_c\to\infty} \lim_{\Delta L/\lambda_q\to\infty} \frac{1}{m} \int_{t_0}^{t} dt( \frac{p\bullet p}{2m} - V_{(q)} - V_{qu(\mathrm{n}_0)} - (V_{qu(\mathrm{n})} - V_{qu(\mathrm{n}_0)} )))\}$$

$$= \lim_{\Delta L/\lambda_c\to\infty} \lim_{\Delta L/\lambda q\to\infty} \frac{1}{m} \nabla_q \{ \int_{t_0}^{t} dt\, ( \frac{p\bullet p}{2m} - V_{(q)} - \lim_{\Theta\to 0} (V_{qu(\mathrm{n})} - V_{qu(\mathrm{n}_0)} )))\} = \frac{p_{cl}}{m} + \frac{\delta p}{m} \cong \frac{p_{cl}}{m} \quad (15)$$

where , given that $\lim_{q\to\infty} -\nabla_q V_{qu} = 0$, it has been posed

$$\lim_{\Delta L/\lambda_c\to\infty} \lim_{\Delta L/\lambda_q\to\infty} V_{qu(\mathrm{n}_0)} = 0, \text{ where}$$

$$\dot{p}_{cl} = \lim_{\Delta L/\lambda_c\to\infty} \lim_{\Delta L/\lambda_q\to\infty} -\nabla_q V_{(q)}, \quad (16)$$

is the classical force and where $\delta p$ is a small fluctuation of the momentum since the variance of the quantum potential fluctuations (i.e., $V_{qu(\mathrm{n})} - V_{qu(\mathrm{n}_0)}$) tends to zero for vanishing $\Theta$ (for the convergence to the deterministic limit warranted by (2)).

By (15) we can see that in the large scale limit the quantum action $S$ that reads

$$\lim_{\Delta L/\lambda_c\to\infty} \lim_{\Delta L/\lambda_q\to\infty} S = \int_{t_0}^{t} dt\, ( \frac{p\bullet p}{2m} - V_{(q)} ) + \delta S_{(V_{qu(\mathrm{n})} - V_{qu(\mathrm{n}_0)})} = S_{cl} + \delta S \quad (17)$$

converges to the classical value $S_{cl}$ plus a small fluctuation and that the noise variance reads

$$\lim_{\Delta L/\lambda_c\to 0} \lim_{\Theta\to 0} < \eta_{(q_\alpha, t)}, \eta_{(q_\alpha + \lambda, t+\tau)} > = \underline{\mu}\, \delta_{\alpha\beta} \frac{2k\Theta}{\lambda_c} \delta(\lambda)\delta(\tau) \quad (18)$$

## 2.2 The features of the large scale classical limit

The stochastic classical dynamics defined by (15-16) is obtained by neglecting the quantum potential interaction that at large distance reads

$$. \lim_{\Delta L/\lambda_c \to \infty} \lim_{\Delta L/\lambda_q \to \infty} V_{qu(n_0)} = 0 << V_{qu(n)} - V_{qu(n_0)}$$

This operation that in absence of noise is not mathematically correct, in presence of noise does not alter the outputs of the equations since it is much smaller than the fluctuations of the quantum potential itself $V_{qu(n)} - V_{qu(n_0)}$.

Obviously, the effect of the quantum potential fluctuations $V_{qu(n)} - V_{qu(n_0)}$ on the dynamics of the system is not equal to the effect of their mean value to which $V_{qu(n_0)}$ may be close. The stochastic sequence of inputs of quantum potential do not lead to any coherent reconstruction of superposition of state whether or not we take into account the very small deterministic quantum potential $V_{qu(n_0)}$.

As a consequence of this selective efficacy of quantum potential interaction that acts on small scale dynamics while it becomes ineffective on large ones, the stochastic large scale mechanics in SQHA maintains its local behavior until the resolution size $\Delta\Omega$ is larger than the quantum coherence length $\lambda_c$.

Moreover, higher is the amplitude of the noise $\Theta$, higher is the attainable degree of spatial precision within the classical (local) limit. On the other hand higher is the amplitude of noise higher are the fluctuations of classical observable such as the variance of energy measurements.

This mutual conflicting effect is basically the same of that one of the conjugated variables leading to the Heisenberg's principle of uncertainty. To verify this, let's evaluate the uncertainty relation between the time interval $\Delta t$ and the variance in the energy measurement on a punctual particle of mass m in the classical mechanical limit of the SQHA.

If on distances smaller than $\lambda_c$ any system behave as a wave so that any its subparts cannot be perturbed without disturbing all the system, it follows that the independence between the measuring apparatus and the measured system requires that they must be far apart more than $\lambda_c$ and hence for the finite speed of propagation of interactions and information the measure process must last longer than the time $\tau = \frac{\lambda_c}{c}$ (see figure 5).

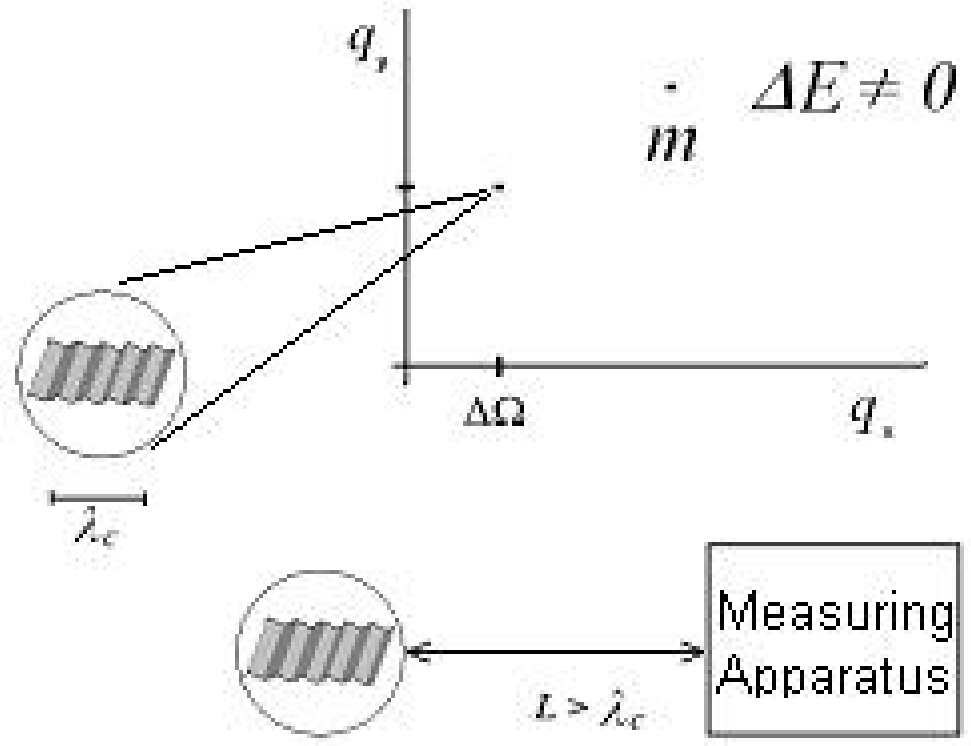

Figure 5. The coarse-grained classical view in a fluctuating quantum environment

Moreover, given the presence of the Gaussian noise we have that the mean value of the energy fluctuation is $\Delta E_{(\Theta)} = k\Theta$.

Thence, for the classical case ($mc^2 >> k\Theta$) for a particle of mass m leads to an energy variance $\Delta E$ that reads

$$\Delta E \approx (<(mc^2 + \Delta E_{(\Theta)})^2 - (mc^2)^2>)^{1/2} \cong (<(mc^2)^2 + 2\Delta E_{(\Theta)} - (mc^2)^2>)^{1/2}$$
$$\cong (2mc^2 < \Delta E_{(\Theta)} >)^{1/2} \cong (2mc^2 k\Theta)^{1/2} \quad (19)$$

from which it follows that

$$\Delta E \Delta t > \Delta E \Delta \tau = \frac{(2mc^2 k\Theta)^{1/2} \lambda_c}{c} = \propto \hbar, \quad (20)$$

therefore if we use the freedom to choose the proportionality constant to introduce into the SQHA model to the physical information of the uncertainty principle $\Delta E \Delta t > \frac{h}{2}$, we obtain that $\propto = \pi$. It is worth noting that the product $\Delta E \Delta \tau$ is constant since the growing of the energy variance with the square root of $\Theta$ $\Delta E \approx (2mc^2 k\Theta)^{1/2}$ is exactly compensated the decrease of the minimum time of measurement

$$\tau \propto \frac{\hbar}{(2mc^2 k\Theta)^{1/2}} \quad (21)$$

The same result is achieved if we derive the uncertainty relations between the position and momentum of a particle of mass m. If we measure the spatial position of a particle with a precision of $\Delta L > \lambda_c$ so that we do not perturb its quantum wave function (that extends its on a smaller spatial domain) the variance $\Delta p$ of the modulus of its relativistic momentum $(p^\mu p_\mu)^{1/2} = mc$ due to the vacuum fluctuations reads

$$\Delta p \approx (<(mc + \frac{\Delta E_{(\Theta)}}{c})^2 - (mc)^2>)^{1/2} \cong (<(mc)^2 + 2m\Delta E_{(\Theta)} - (mc)^2>)^{1/2}$$
$$\cong (2m < \Delta E_{(\Theta)} >)^{1/2} \cong (2mk\Theta)^{1/2} \quad (22)$$

Leading to the uncertainty relation\

$$\Delta L \Delta p > \lambda_c (2mk\Theta)^{1/2} = \frac{h}{2} \quad (23)$$

If we require to measure the spatial position with a precision $\Delta L < \lambda_c$, we have to perturb the particle quantum state.

Due to the spatial confinement of the wave function, an increase of the quantum potential energy is generated so that the final particle momentum gets a variance $\Delta p$ higher than (22) and still satisfying the uncertainty relations steaming by the property of the Fourier transform relations.

(22) holds as far as we remain in the classical approach and we do not explore the space into domains whose length is smaller than $\lambda_c$ so that momentum variance is given just by the environmental fluctuations.

## 3. DISCUSSION

The SQHA model shows that for $\Theta$ that goes to zero ( i.e., $\lambda_c \to \infty$ and the standard quantum mechanics is realized) the measuring time goes to infinity and the energy fluctuation $\Delta E$ goes to zero (we have a perfect overall quantum system with exactly defined energy levels).
This makes clear that in a perfect quantum universe the measuring process is endless (i.e., not possible) confirming that the classical behavior is needed to the definition of the quantum mechanics based on the measuring process.
The SQHA approach shows that the logical contradiction between the non-local quantum interaction and the finite transmission speed of interactions comes by two singularities of the standard approach: the infinite velocity of light in the non relativistic limit and the infinite value of the quantum coherence length of the deterministic limit of standard quantum mechanics.
This conceptual incongruity ends in the frame of the SQHA model where quantum coherence length is finite and the systematic passage to the classical mechanics is allowed. On macroscopic scale with the range of non-local interaction equal to $\lambda_c$ (i.e., local classical description) the minimum uncertainty principle is compatible with the relativistic requirement of finite speed of light and information transmission. It is interesting to note that the uncertainty relations still apply to the spatial localization of quantum state due to the presence of fluctuations.
Finally, it is worth mentioning that the Heisenberg indetermination principle in the SQHA is achieved by the fact that the minimum duration of the measurement process multiplied by variance of the energy fluctuation leads to a constant. This can happen since the energy variance going like $\Delta E \approx (2mc^2 k\Theta)^{1/2}$ exactly compensates the $\Theta$-dependence of the minimum time of measurement $\tau \propto \frac{\hbar}{(2mc^2 k\Theta)^{1/2}}$. The same compensation happens for the position-momentum uncertainty relation of a particle of mass m.
If we consider the SQHA model as the general one, in which the standard quantum mechanics represents the deterministic limit, and that the minimum uncertainty relations comes from a property of the general SQHA theory, it follows that the uncertainty postulate can be re-visited as deriving by the finite transmission speed of interactions and the existence of a finite non-local quantum distance in a fluctuating environment.

## 4. CONCLUSION

The stochastic generalization of the QHA shows that the local classical description, in which it is possible to divide a system into independent sub-parts (so that we can improve the precision of local variables) has a physical limit given by the distance on which the non-local quantum interaction takes place.
Given that below the length on which the quantum non-locality sets-in, any subpart of a system cannot collect and extract statistically independent information on the remaining one, the measuring process cannot be performed in a time shorter than that one needed to the interactions and information to be transmitted beyond such a distance.

The localization of the quantum states in the SQHA model (achieved as a consequence of fluctuations) obeys to the uncertainty relations.
The existence of the Heisenberg indetermination principle into the SQHA dynamics can be derived by the fact that the minimum duration of the measurement process multiplied by variance of the energy fluctuation results constant.
In the frame of this approach the minimum uncertainty postulate derives by requiring that the finite transmission speed of information is compatible with the non-local character of the quantum mechanics.

NOMENCLATURE

| | |
|---|---|
| n : squared wave function modulus | $l^{-3}$ |
| $S$ : action of the system | $m^{-1}\ l^{-2}\ t$ |
| $m$ : mass of structureless particles | m |
| $\hbar$ : Plank's constant | $m\ l^{2}\ t^{-1}$ |
| $c$ : light speed | $l\ t^{-1}$ |
| $k$ : Boltzmann's constant | $m\ l^{2}\ t^{-2}\ °K^{-1}$ |
| $\Theta$ : Noise amplitude | °K |
| $H$ : Hamiltonian of the system | $m\ l^{2}\ t^{-2}$ |
| $V$ : potential energy | $m\ l^{2}\ t^{-2}$ |
| $V_{qu}$ : quantum potential energy | $m\ l^{2}\ t^{-2}$ |
| $\eta$ : Gaussian noise of WFMS | $l^{-3}\ t^{-1}$ |
| $\lambda_c$ : correlation length of squared wave function modulus fluctuations | l |
| $\lambda_L$ : range of interaction of non-local quantum interaction | l |
| $G(\lambda)$ : dimensionless correlation function (shape) of WFMS fluctuations | pure number |
| $\underline{\mu}$: WFMS mobility form factor | $m^{-1}\ t\ \ l^{-6}$ |
| $\mu$ = WFMS mobility constant | $m^{-1}\ t$ |

## COMPETING INTERESTS

The Author declares that no competing interests exist.

## AUTHORS' CONTRIBUTIONS

The Author designed the study, performed the mathematical analysis, wrote the discussion and managed the literature searches. The author read and approved the final manuscript